\documentclass[aip,jcp,reprint]{revtex4-1}
\usepackage{graphicx,dcolumn,xcolor,microtype,multirow,amscd,amsmath,amssymb,amsfonts,physics,wrapfig,xspace}
\usepackage[version=4]{mhchem}
\usepackage{chemfig}
\usepackage{subcaption}
\usepackage{mathtools}

\usepackage{siunitx}
\DeclareSIUnit[number-unit-product = {\,}]
\cal{cal}
\DeclareSIUnit\kcal{\kilo\cal}

\DeclareSIUnit\debye{D}

\usepackage{hyperref}
\hypersetup{
    colorlinks,
    linkcolor={red!50!black},
    citecolor={red!70!black},
    urlcolor={red!80!black}
}

\newcommand{\ie}{\textit{i.e.}}

\newcommand{\alert}[1]{\textcolor{black}{#1}}
\usepackage[normalem]{ulem}

\newcommand{\SupMat}{\textcolor{blue}{supporting information}\xspace}

\newcommand{\mc}{\multicolumn}
\newcommand{\fnm}{\footnotemark}
\newcommand{\fnt}{\footnotetext}

\newcommand{\br}{\boldsymbol{r}}

\newcommand{\etaopt}{\eta_{\text{opt}}}
\DeclareMathOperator*{\argmin}{\arg\min}

\newcommand{\hW}{\Hat{W}}
\newcommand{\hw}{\Hat{w}}
\newcommand{\hH}{\Hat{H}}
\newcommand{\hHcap}{\Hat{H}(\eta)}

\newcommand{\cmel}[3]{\mel*{#1}{#2}{#3}_\text{c}}
\newcommand{\cbraket}[2]{\braket*{#1}{#2}_\text{c}}

\newcommand{\Evar}{E_\text{var}}
\newcommand{\Eexfci}{E_\text{exFCI}}
\newcommand{\EexfciA}{E_\text{exFCI}^\text{A}}
\newcommand{\EexfciN}{E_\text{exFCI}^\text{N}}
\newcommand{\Esta}{E_\text{sta}}
\newcommand{\Ept}{E_\text{PT2}}

\newcommand{\Eapt}{E_\text{aPT2}}
\newcommand{\Psivar}{\Psi_\text{var}}
\newcommand{\Psista}{\Psi_\text{sta}}

\newcommand{\ii}{\mathrm{i}}

\newcommand{\LCPQ}{Laboratoire de Chimie et Physique Quantiques (UMR 5626), Universit\'e de Toulouse, CNRS, UPS, 31062 Toulouse, France}

\begin{document}

\title{Selected Configuration Interaction for Resonances}

\author{Yann \surname{Damour}}
	\email{yann.damour@irsamc.ups-tlse.fr}
	\affiliation{\LCPQ}
\author{Anthony \surname{Scemama}}
	\affiliation{\LCPQ}
\author{F\'abris \surname{Kossoski}}
	\email{fabris.kossoski@irsamc.ups-tlse.fr}
	\affiliation{\LCPQ}
\author{Pierre-Fran\c{c}ois \surname{Loos}}
	\email{loos@irsamc.ups-tlse.fr}
	\affiliation{\LCPQ}

\begin{abstract}
Electronic resonances are metastable states that can decay by electron loss.
They are ubiquitous across various fields of science, such as chemistry, physics, and biology.
However, current theoretical and computational models for resonances
cannot yet rival the level of accuracy achieved by bound-state methodologies.
Here, we generalize selected configuration interaction (SCI) to treat resonances using the complex absorbing potential (CAP) technique.
By modifying the selection procedure and the extrapolation protocol of standard SCI, the resulting CAP-SCI method
yields resonance positions and widths of full configuration interaction quality.
Initial results for the shape resonances of \ce{N2-} and \ce{CO-} reveal the important effect of high-order correlation,
which shifts the values obtained with CAP-augmented equation-of-motion coupled-cluster with singles and doubles by more than \SI{0.1}{\eV}.
The present CAP-SCI approach represents a cornerstone in the development of highly-accurate methodologies for resonances.
\bigskip
\begin{center}
       \boxed{\includegraphics[width=0.4\linewidth]{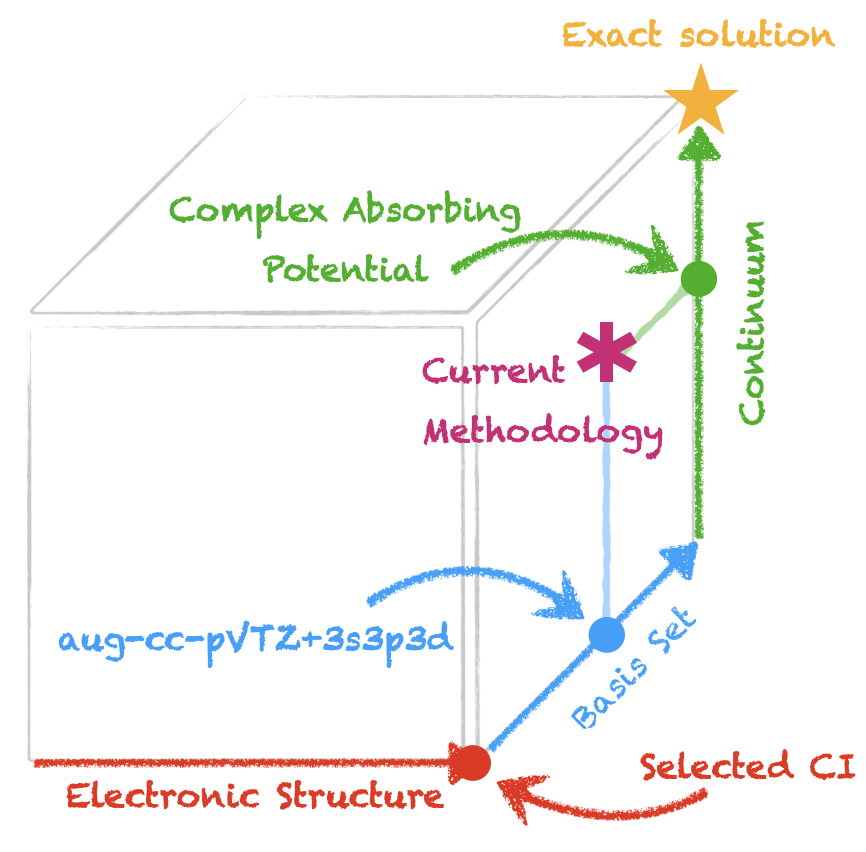}}
\end{center}
\end{abstract}

\maketitle


The electronic spectrum of molecular systems contains continuum (unbound) states in addition to the usual discrete (bound) states.
Embedded in the continuum, one can find electronic resonances,
metastable states that can decay by losing one electron. \cite{Klaiman_2012}
In contrast to the real energies of bound states,
resonances have a complex-valued energy
\begin{equation}
E = E_R - \ii \Gamma/2,
\label{eq:energy}
\end{equation}
with resonance position $E_R$ and resonance width $\Gamma$ (related to its lifetime against autoionization $\hbar/\Gamma$).

As a widespread type of resonance, we cite temporary anions. \cite{Simons_2023,Clarke_2024}
They can be formed by electron attachment to a molecule or by photoexcitation of a bound anion.
Temporary anions play important roles in various fields of chemistry and physics.
To name a few, they are involved in the DNA damage induced by ionizing radiation, \cite{Boudaiffa_2000,Alizadeh_2015} in the bioactivity of some classes of radiosensitizers, \cite{Sedmidubska_2024}
in the chemistry of interstellar medium, \cite{Wu_2024} and in different technologies for nanofabrication. \cite{Arumainayagam_2010,Thorman_2015}
Other types of resonances include multiply charged anions,
core-excited, core-ionized, and superexcited states,
and Stark resonances (formed by exposing a molecule to a strong electric field). \cite{Jagau_2017,Jagau_2022}

To describe molecular resonances, one must solve a quantum many-body problem (just as for bound states),
while accounting for its coupling with the continuum (which is not an issue for bound states).
The problem of electronic correlation in the continuum represents a tremendous challenge for theory.

Scattering methodologies can formally produce a complete description of resonances and their embedding continuum. \cite{Domcke_1991,Gianturco_2013}
However, these methods are usually coupled with more approximate treatments for the electronic correlation,
which currently limits their ability to produce reliably accurate resonance energies. \cite{Schneider_1988,Tennyson_2010,DaCosta_2015}

Alternatively, one can resort to adapted quantum chemistry methodologies.
They retain the stationary-like picture of a resonance (as in bound-state quantum chemistry),
while the effect of the continuum is accounted for implicitly.
To do so, one can stay in a Hermitian formulation by employing stabilization techniques. \cite{Hazi_1970,Taylor_1970}
Instead, one can shift to a non-Hermitian formulation of quantum mechanics. \cite{Moiseyev_2011,Jagau_2017,Jagau_2022}
In this case, the Hamiltonian becomes complex-valued and non-Hermitian.
As a consequence, the resonance directly emerges as an eigenstate of this modified Hamiltonian, having a complex energy.
In this class of complex variable methods, we find complex scaling, \cite{Balslev_1971,Moiseyev_1998} complex basis functions, \cite{McCurdy_1978,White_2015} and complex absorbing potential (CAP). \cite{Jolicard_1985,Riss_1993,Sommerfeld_1998,Ghosh_2012,Zuev_2014,Jagau_2014a,Jagau_2014b,Sommerfeld_2015,Gyamfi_2024}

In the complete basis set limit and in the full configuration interaction (FCI) limit, the continuum techniques mentioned above yield the exact resonance energy.
In practice, finite basis sets and approximate electronic structure theories must be employed.
Various continuum techniques have been combined with many different levels of theory, giving rise to a wide range of methods.
In particular, the CAP is one of the most widely used techniques for studying resonances.
It has been combined with equation-of-motion electron-attached coupled-cluster with singles and doubles (EOM-EA-CCSD), \cite{Ghosh_2012,Ghosh_2013,Zuev_2014}
Fock-space multireference coupled-cluster, \cite{Sajeev_2005a,Sajeev_2005b}
coupled-cluster with perturbative triples, \cite{Jagau_2018,Jana_2021}
algebraic diagrammatic construction, \cite{Santra_2002,Feuerbacher_2003,Belogolova_2021}
symmetry-adapted-cluster configuration interaction, \cite{Ehara_2012}
extended multiconfigurational quasidegenarate perturbation theory of the second order, \cite{Kunitsa_2017,Phung_2020}
multireference configuration interaction, \cite{Sommerfeld_1998,Sommerfeld_2001}
and density-functional theory. \cite{Zhou_2012}

Despite significant theoretical advances in recent years, \cite{Jagau_2017,Jagau_2022}
the most accurate existing methods for molecular resonances
cannot rival the level of accuracy that has been achieved for bound states. \cite{Loos_2020d}
Strikingly, no methodology can systematically approach the FCI limit for resonances.
As a consequence, highly-accurate resonance energies, that is, with uncertainties below \SI{1}{kcal/mol} or \SI{0.043}{\eV}, remain out of reach.

To close this gap between methodologies for bound states and resonances, here we combine selected configuration interaction (SCI)
\cite{Bender_1969,Huron_1973,Buenker_1974,Evangelisti_1983,Harrison_1991,Angeli_2001c,Giner_2013,Holmes_2016,Schriber_2016,Tubman_2016,Liu_2016b,Sharma_2017,Coe_2018,Garniron_2019,Zhang_2020}
with the CAP technique.
This choice is motivated by the ability of SCI to systematically approach the FCI limit for bound states. \cite{Eriksen_2020a,Loos_2020i,Eriksen_2021,Larsson_2022,Damour_2023}
It is able to provide highly-accurate excitation energies,
allowing a faithful benchmark of more approximate methods. \cite{Holmes_2017,Schriber_2017,Chien_2018,Loos_2018a,Veril_2021,Prentice_2019,Zhang_2020,Coe_2022}
With the present CAP-SCI methodology, similarly accurate resonance energies can be envisioned.
As a first application of this novel methodology, we address the emblematic shape resonances of \ce{N2-} and \ce{CO-}.
Unless otherwise stated, atomic units are used throughout.



To absorb the oscillating tail of the resonance wave function and render it square-integrable,
a CAP $\hW$ of strength $\eta > 0$ is added to the physical $N$-body electronic Hamiltonian $\hH$,
yielding a perturbed and $\eta$-dependent Hamiltonian
\begin{equation}
\hHcap = \hH - \ii \eta \hW
\end{equation}
with
\begin{equation}
\hW = \sum_{k=1}^N \hw_k
\end{equation}
The one-body potential $\hw$ can be chosen from various functional forms but the most widely used form remains the quadratic potential
\begin{equation}
\hw_{\alpha_k} =
\begin{cases}
\qty(\abs{\alpha_k} - \alpha_0)^2 & \text{if $\abs{\alpha_k} > \alpha_0$}
\\
0 & \text{otherwise}
\end{cases}
\end{equation}
where $\hw_k  = \hw_{x_k} + \hw_{y_k} + \hw_{z_k}$, $\alpha_k \in \{x_k,y_k,z_k\}$ is the coordinate of the $k$th electron, and $\alpha_0 \in \{x_0,y_0,z_0\}$ defines the CAP onset.
It is important to notice that $\hW$ is a symmetric operator.
As a consequence, while $\hH$ is Hermitian, $\hHcap$ is a complex symmetric operator, that is, $\hHcap = \hHcap^T$, leading to complex-valued eigenvalues and eigenvectors, \alert{the left and right eigenvectors being related by transposition.}

Because of the non-Hermitian nature of $\hHcap$, the usual variational principle of the real-valued energy has to be replaced by a stationary principle for the complex-valued energy based on the c-product \cite{Moiseyev_1978}
\begin{equation}
E(\eta) = \cmel{\Psi}{\hHcap}{\Psi}
\label{eq:stationary_principle}
\end{equation}
where $\Psi$ is a c-normalized trial wave function, i.e., $\cbraket{\Psi}{\Psi} = 1$, and
\begin{equation}
\cbraket{f}{g} = \int f(\br) g(\br) \dd{\br}
\label{eq:c-prod}
\end{equation}

In a complete basis set, the resonance position and width can be extracted by computing the energy as $\eta \to 0^+$.
However, in a finite basis set, one must find a non-zero and optimal $\etaopt$ that balances out the
error stemming from the CAP (which increases with $\eta$) and the basis set incompleteness error (which decreases with $\eta$).
As shown by Riss and Meyer, \cite{Riss_1993} this can be achieved by minimizing the energy velocity
\begin{equation}\label{eq:eta_opt}
\etaopt = \argmin_\eta \abs{\eta \dv{E(\eta)}{\eta}}
\end{equation}
The practical procedure to find $\etaopt$ generally consists of computing
$\eta$ ``trajectories'', \ie, the evolution of $E(\eta)$ as a function of $\eta$,
and then look for the minimum of the energy velocity defined in Eq.~\eqref{eq:eta_opt} along the trajectory.

To reduce the dependence of the zeroth-order energy $E(\etaopt)$ on the CAP parameters,
it is a common practice to compute the first-order corrected energy \cite{Jagau_2014a,Jagau_2014b}
\begin{equation} \label{eq:E_tilde}
	\Tilde{E}(\etaopt) = E(\etaopt) - \etaopt \eval{\dv{E(\eta)}{\eta}}_{\eta = \etaopt}
\end{equation}
where the derivative is computed as \cite{Jagau_2014a}
\begin{equation}
\eval{\dv{E(\eta)}{\eta}}_{\eta=\etaopt} = - \ii \Tr[\hat{\gamma}(\etaopt) \hw]
\end{equation}
with $\hat{\gamma}(\eta)$ the one-particle density operator.


Here, we rely on the \textit{``Configuration Interaction using a Perturbative Selection made Iteratively''} (CIPSI) method, \cite{Huron_1973,Giner_2013,Giner_2015,Garniron_2017,Garniron_2018,Garniron_2019}
one of the numerous variants belonging to the SCI family. \cite{Schriber_2016,Tubman_2016,Tubman_2018,Tubman_2020,Holmes_2016,Holmes_2017,Sharma_2017,Chien_2018,Yao_2020,Yao_2021,Larsson_2022,Liu_2014,Liu_2016,Lei_2017,Zhang_2020,Zhang_2021}
The CIPSI algorithm is well documented in the literature \cite{Garniron_2018,Garniron_2019} and, thus, we only summarize below the main steps to highlight its generalization to a non-Hermitian, complex-valued framework.

In the standard (real-valued) Hermitian case, the variational wave function associated with the electronic state of interest is written as
\begin{equation} \label{eq:Psivar}
\ket{\Psivar} = \sum_{I \in \mathcal{I}} c_I \ket{I}
\end{equation}
where the $\ket{I}$'s are Slater determinants belonging to the so-called internal (or variational) space $\mathcal{I}$.
The real-valued variational energy associated to this wave function can be computed as
\begin{equation} \label{eq:Evar}
\Evar = \mel{\Psivar}{\hH}{\Psivar}
\end{equation}
where $\Psivar$ is chosen normalized, i.e., $\braket{\Psivar}{\Psivar}=1$.
The variational energy \eqref{eq:Evar} and the real-valued coefficients $c_I$ appearing in Eq.~\eqref{eq:Psivar}
are obtained by diagonalization of the CI matrix with elements $\mel*{I}{\hH}{J}$ using the Davidson algorithm. \cite{Davidson_1975,Garniron_2019}

To complement $\Evar$, a second-order perturbative correction, $\Ept$, is usually added to it.
Although non-variational, the resulting energy, $\Evar+\Ept$, is a more faithful approximation of the FCI energy.
By employing the Epstein-Nesbet partitioning, the expression of the second-order energy is expressed as follows:
\begin{equation} \label{eq:Ept}
\Ept = \sum_{\alpha} e_\alpha^{(2)} = \sum_\alpha \frac{\mel*{\alpha}{\hH}{\Psivar}^2}{\Evar - \mel*{\alpha}{\hH}{\alpha}}
\end{equation}
where the $\ket{\alpha}$'s are Slater determinants belonging to the external (or perturbative) space $\mathcal{A}$,
such that $\alpha \notin \mathcal{I}$ and $\mel*{I}{\hH}{\alpha} \neq 0$.
In our implementation, this perturbative correction is computed using an efficient semistochastic algorithm. \cite{Garniron_2017}

In SCI algorithms like CIPSI, the internal space $\mathcal{I}$ expands iteratively through the inclusion of determinants from the external space $\mathcal{A}$.
These determinants are chosen based on their contribution to the second-order perturbative energy, $e_\alpha^{(2)}$ [see Eq.~\eqref{eq:Ept}].
In practice, we double the size of the variational space at each iteration by incorporating the determinants with the largest values of $\abs*{e_\alpha^{(2)}}$.
Additional determinants are added in order to generate pure spin states. \cite{Chilkuri_2021}

For large enough variational spaces, there exists a rigorous linear relationship between $\Evar$ and $\Ept$. \cite{Burton_2024}
When this linear regime is reached, it is thus possible to extrapolate $\Evar$ to the limit where $\Ept \to 0$, which effectively corresponds to the FCI limit.
More information about the theoretical foundation of the extrapolation procedure can be found in Ref.~\onlinecite{Burton_2024}.


In the case of a CAP-augmented Hamiltonian $\hHcap$, key changes arise.
The stationary wave function associated with the state of interest reads
\begin{equation}
\ket{\Psista(\eta)} = \sum_{I \in \mathcal{I}} c_I(\eta) \ket{I}
\end{equation}
where the CI coefficients $c_I(\eta)$ are now complex-valued.
Moreover, based on the stationary principle defined in Eq.~\ref{eq:stationary_principle},
the expression of the complex-valued stationary energy is
\begin{equation} \label{eq:Esta}
\Esta(\eta) = \cmel{\Psista(\eta)}{\hHcap}{\Psista(\eta)}
\end{equation}
\alert{with the c-normalization condition $\cbraket{\Psista(\eta)}{\Psista(\eta)} = 1$, as defined in Eq.~\eqref{eq:c-prod}}.
Likewise, the complex-valued second-order perturbative energy reads
\begin{equation} \label{eq:Ept_c}
\Ept(\eta) = \sum_{\alpha} e_\alpha^{(2)}(\eta) = \sum_\alpha \frac{\cmel{\alpha}{\hHcap}{\Psista(\eta)}^2}{\Esta(\eta) - \cmel{\alpha}{\hHcap}{\alpha}}
\end{equation}
As in the Hermitian case, the stationary energy \eqref{eq:Esta} and the coefficients $c_I(\eta)$ are obtained by diagonalization of the CI matrix with elements $\cmel{I}{\hH(\eta)}{J}$
using the Davidson algorithm adapted for symmetric complex matrices, \cite{Hirao_1982}
while the complex-valued perturbative correction \eqref{eq:Ept_c}
is computed with a straightforward generalization of the semistochastic algorithm developed in Ref.~\onlinecite{Garniron_2017}.
\alert{Note that we did not encounter self-orthogonality issues (the c-norm may be zero for non-zero functions) during the iterative Davidson diagonalization process.}

In a complex-valued setup, the selection procedure is more intricate as one must select determinants that contribute to the real and imaginary parts of the stationary energy.
The most natural and democratic way consists of selecting determinants $\ket{\alpha}$'s based on the largest $\abs*{e_\alpha^{(2)}(\eta)}$ values, as in the Hermitian case (see above).
However, as we shall see below, it is also possible to accelerate the convergence of either the real or imaginary part of the energy by employing $\Re[e_\alpha^{(2)}(\eta)]$ or $\Im[e_\alpha^{(2)}(\eta)]$ as a selection criterion, respectively.

To produce the final FCI estimates of the resonance position and width,
we carefully monitor the behavior of the stationary energy $\Esta(\eta)$
as a function of the perturbative correction $\Ept(\eta)$.
For sufficiently large stationary wave functions, we observe that $\Re[\Esta(\eta)]$ and \alert{$\Im[\Esta(\eta)]$}
behave linearly with respect to $\Re[\Ept(\eta)]$ and $\Im[\Ept(\eta)]$, respectively.

We further notice that, for a sufficiently large stationary space, because the real components are always much larger than their imaginary analogs and $\Re[\Esta(\eta)] \ll \Re[\cmel{\alpha}{\hHcap}{\alpha}]$, we have $\Re[e_\alpha^{(2)}(\eta)] < 0$ for any external determinant $\alpha$.
In other words, the real part of Eq.~\ref{eq:Ept_c} is a sum of negative terms only and, thus, $\Re[\Ept(\eta)]$ approaches zero from below.
In contrast, the sign of $\Im[e_\alpha^{(2)}(\eta)]$ can be either positive or negative.
This has two major consequences: i) it is not possible to anticipate how $\Im[\Ept(\eta)]$ approaches zero, and ii) the condition $\Im[\Ept(\eta)] = 0$ does not necessarily imply that the FCI limit has been reached.

To address this issue, we introduce the ``absolute value'' version of the second-order energy
\begin{equation}
\Eapt(\eta) = \sum_{\alpha} \abs{\Re[e_\alpha^{(2)}(\eta)]} + \ii \sum_{\alpha} \abs{\Im[e_\alpha^{(2)}(\eta)]}
\end{equation}
such that
\begin{equation}
\Im[\Eapt(\eta)] = \sum_{\alpha} \abs{\Im[e_\alpha^{(2)}(\eta)]}
\end{equation}
for which the condition $\Im[\Eapt(\eta)] = 0$ is fulfilled only when the FCI limit has been attained.
The FCI estimates $\Eexfci$ are thus obtained through independent linear extrapolations of $\Re[\Esta(\eta)]$ as $\Re[\Ept(\eta)] \to 0$ and of $\Im[\Esta(\eta)]$ as $\Im[\Eapt(\eta)] \to 0$.


As illustrative examples, we consider the widely studied $^2 \Pi_g$ shape resonance of \ce{N2-} and $^2 \Pi$ shape resonance of \ce{CO-}.
Shown in Table \ref{tab:comp_det} are the geometries, basis sets, CAP onsets and strengths,
all taken from Ref.~\onlinecite{Zuev_2014}, which reported CAP-EA-EOM-CCSD resonance energies for these two systems.
We employed the aug-cc-pVTZ basis set with additional 3s3p3d basis functions located at the geometric center of the molecule,
with exponents chosen as described by Zuev \textit{et al.} \cite{Zuev_2014} (reproduced in the \SupMat).
To comply with Ref.~\onlinecite{Zuev_2014}, the frozen-core approximation was not enforced (all electrons were correlated).
To mitigate the unphysical perturbation caused by the CAP, it is common practice to retain solely the virtual-virtual block of the CAP matrix. \cite{Santra_1999}
However, we opted against this strategy because it introduces an artificial dependence of the FCI energies on the choice of orbitals.

\begin{table}
	\caption{Parameters employed for the calculation of the shape resonance of \ce{N2-} and \ce{CO-}:
	bond length (in bohr), CAP onset $(x_0,y_0,z_0)$ (in bohr), and optimal CAP strength $\etaopt$ (in a.u.), all taken from Ref.~\onlinecite{Zuev_2014}.
	The molecular axis is chosen as the $z$ axis and the basis set is aug-cc-pVTZ+3s3p3d for both systems (the additional basis functions are centered at the geometric center of the molecule, located at the origin).}
	\label{tab:comp_det}
	\begin{ruledtabular}
		\begin{tabular}{cccccc}
			System		&	State 		&	Bond length	&	$(x_0,y_0,z_0)$		&	$\etaopt$	&	\\
			\hline
			\ce{N2-}	&	$^2 \Pi_g$	&	$2.0740$	&	$(2.76,2.76,4.88)$	&	0.0015	\\
			\ce{CO-}	&	$^2 \Pi$	&	$2.1316$	&	$(2.76,2.76,4.97)$	&	0.0028	\\
		\end{tabular}
	\end{ruledtabular}
\end{table}

Notice that we adopt the CAP strength $\etaopt$ optimized for CAP-EA-EOM-CCSD. \cite{Zuev_2014}
Here, we do not discuss its optimization within the CAP-SCI methodology,
since our goal is to gauge the convergence of the resonance energy towards the FCI limit for fixed CAP parameters.
This allows us to attribute the differences between our CAP-SCI and the CAP-EA-EOM-CCSD results \cite{Zuev_2014}
exclusively to electronic correlation effects.
Because of that and to shorten the notation, we drop the explicit dependence on $\eta$ from hereon.

All calculations were performed with \textsc{quantum package}, \cite{Garniron_2019} where we implemented the CAP-SCI scheme.
Since \textsc{quantum package} relies on the CIPSI flavor of SCI methods, our particular implementation of CAP-SCI is labeled CAP-CIPSI.

Our calculations relied on a state-following procedure \cite{Butscher_1976}
which we also implemented in \textsc{quantum package}.
At each iteration of the Davidson diagonalization,
we keep track of the state of interest by monitoring the overlap between the input and Ritz vectors.  \cite{Butscher_1976}
This algorithm enables us to carry out CIPSI calculations for a state that is not the lowest-lying of a given symmetry sector.
It proved to be very useful to accelerate the convergence of the present calculations.

Here, we restricted ourselves to real-valued orbitals,
which was driven by their simplicity when compared to the complex-valued alternative.
It allows us to recycle a significant number of functions from the standard real-valued CIPSI algorithm. \cite{Garniron_2019}
Additionally, it reduces both the memory requirement for the two-electron integrals in the orbital basis
and the computational cost in the calculation of quantities derived from these integrals.

Our computational procedure for the anionic state is as follows.
Starting from a given set of orbitals (described below), we perform an extended CAP-CIS calculation. \cite{Maurice_1996}
The resonance state can be easily identified by the occupation of a $\pi^*$ orbital and by an imaginary part of the energy that is not too large (see the \SupMat).
Then, we remove all the determinants having small contributions to this state.
Finally, a single-state CAP-CIPSI calculation is performed using the state-following procedure described above,
until the wave function reaches around \num{4d7} determinants.

Although the FCI energy is unaffected by the underlying set of orbitals employed for constructing Slater determinants,
in practice, the choice of orbitals can strongly influence how fast the FCI estimate is reached.
Four different sets were tested:
restricted Hartree-Fock (HF) orbitals of the neutral species (HFN),
restricted open-shell HF orbitals of the anion species (HFA),
natural orbitals of the neutral species (NON), and natural orbitals of the anion species (NOA).
To compute the natural orbitals, a preliminary CAP-CIPSI calculation was performed
(with HFN orbitals for the neutral and with HFA orbitals for the anion),
which was ended when the wave function contained at least \num{5d6} determinants.
Then, real-valued natural orbitals were computed by diagonalizing the one-electron reduced density matrix at $\eta = 0$.

The FCI estimates of the total energy, $\Re[\Eexfci]$ and $\Im[\Eexfci]$, are computed using four-point linear fits using the largest stationary wave functions.
The extrapolated FCI estimates thus have associated extrapolation errors, which are given in parenthesis.
Results produced by three- and five-point linear regressions are reported in the \SupMat.

To obtain FCI estimates of the resonance positions $E_R$ and widths $\Gamma$,
a CAP-CIPSI calculation was also performed for the neutral species with NON (for the same $\etaopt$ given in Table \ref{tab:comp_det}).
$E_R$ and $\Gamma$ were extracted from the corresponding
differences between the energies of the anion $\EexfciA$ and of the neutral $\EexfciN$:
\begin{subequations}
\begin{align}
\label{eq:E_opt}
E_R & = \Re[\EexfciA - \EexfciN]
\\
\Gamma & = -2 \Im[\EexfciA - \EexfciN]
\end{align}
\end{subequations}
The uncertainties of the resonance parameters are derived from the uncertainties of $\EexfciA$ and $\EexfciN$.
It is important to mention that the CAP has a fairly small effect on the energy of the neutral systems.
Considering the largest wave functions for \ce{N2} and \ce{CO}, had we adopted the CAP-free results,
the zeroth-order resonance positions would be affected by around \SI{0.1}{\meV} and the widths by less than \SI{5}{\meV},
with even smaller effects for the first-order values.



We start by investigating the influence of the choice of orbitals on the convergence of the CAP-CIPSI calculations.
This analysis is done for the $^2 \Pi_g$ shape resonance of \ce{N2-} and for the selection criterion based on $\abs*{e_\alpha^{(2)}}$
(the comparison between the different selection criteria is presented later).

\begin{figure*}
    \centering
    \begin{subfigure}[b]{0.475\textwidth}
        \centering
        \includegraphics[width=\textwidth]{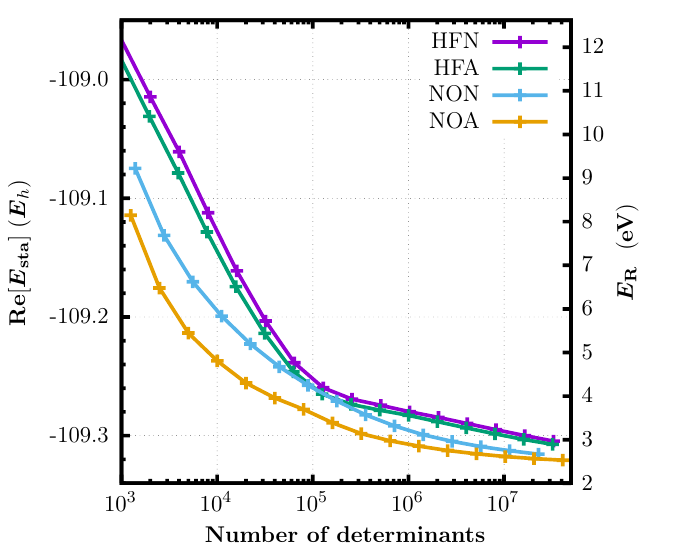}
        \caption{}
        \label{fig:mos_re_e_f_ndet}
    \end{subfigure}
    \hfill
    \begin{subfigure}[b]{0.475\textwidth}
        \centering
        \includegraphics[width=\textwidth]{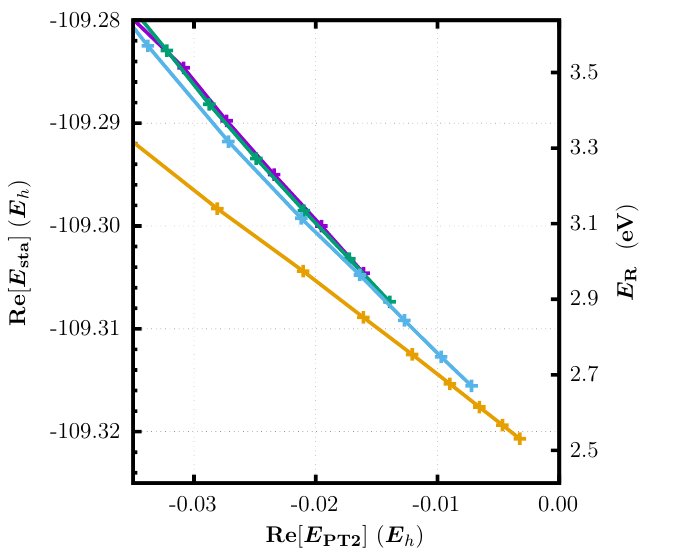}
        \caption{}
        \label{fig:mos_re_e_f_pt2}
    \end{subfigure}
    \vskip\baselineskip
    \begin{subfigure}[b]{0.475\textwidth}
        \centering
        \includegraphics[width=\textwidth]{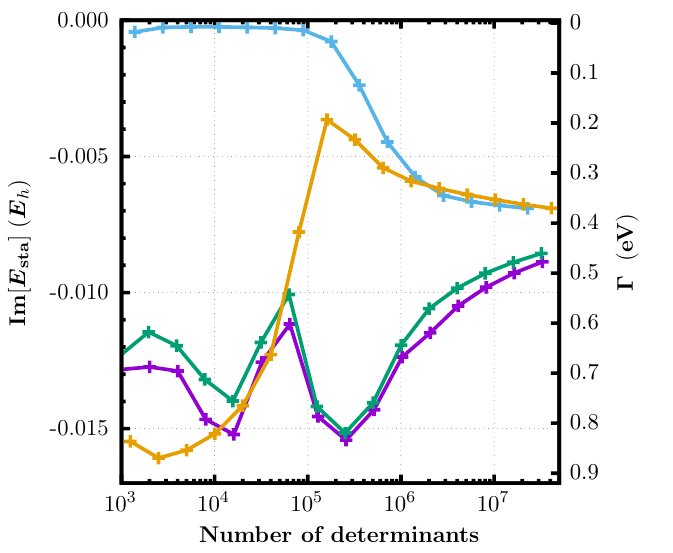}
        \caption{}
        \label{fig:mos_im_e_f_ndet}
    \end{subfigure}
    \hfill
    \begin{subfigure}[b]{0.475\textwidth}
        \centering
        \includegraphics[width=\textwidth]{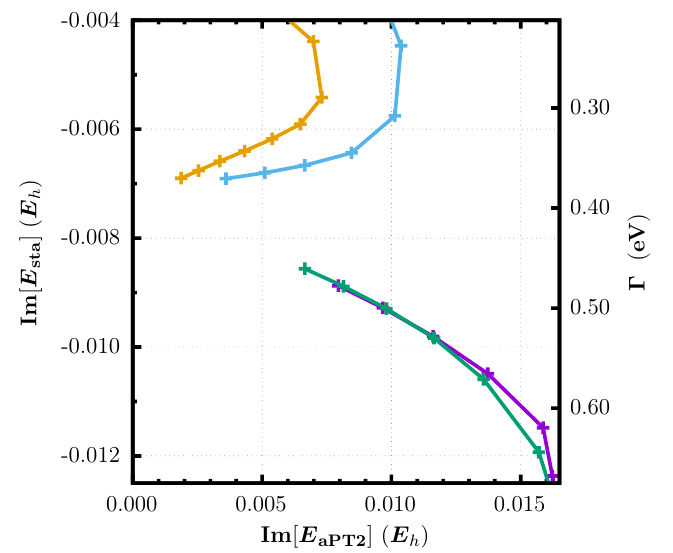}
        \caption{}
        \label{fig:mos_im_e_f_apt2}
    \end{subfigure}
	\caption{Evolution of the real and imaginary parts of the CAP-CIPSI energy, $\Re[\Esta]$ and $\Im[\Esta]$,
	as functions of the number of determinants [panels (a) and (c)] and as functions of their corresponding second-order energy corrections, $\Re[\Ept]$ and $\Im[\Eapt]$, [panels (b) and (d)] for the $^2\Pi_g$ shape resonance of \ce{N2-} with the parameters of Table \ref{tab:comp_det}.
	Four different sets of orbitals are considered: HFN, HFA, NON, and NOA (see main text for more details).
	Resonance position $E_R$ and width $\Gamma$ are obtained for the fixed extrapolated energy of the neutral system.
	}
    \label{fig:mos}
\end{figure*}

The evolution of $\Re[\Esta]$ as a function of the number of determinants in the stationary space is illustrated
in Fig.~\ref{fig:mos_re_e_f_ndet} for each set of orbitals.
All curves display a smooth convergence from above, similar to what is usually observed in standard CIPSI calculations for bound states.
Natural orbitals provide a faster convergence than HF orbitals,
whereas orbitals optimized for the anion are preferable with respect to orbitals obtained for the neutral species.
For example, the energy obtained with HFN and \num{3.2e7} determinants,
is similar to the energy that one would reach with HFA and \num{1.6e7} determinants,
which in turn is reached with as few as \num{2.9e6} and \num{6.2e5} determinants with NON and NOA, respectively.

To obtain $\Eexfci$ values, we look into the evolution of $\Re[\Esta]$ as a function of $\Re[\Ept]$, which is shown in Fig.~\ref{fig:mos_re_e_f_pt2}.
All stationary energies enter into a linear regime when $\Re[\Ept]$ reaches \SI{-0.03}{\hartree}.
This occurs between \num{1e6} and \num{2e6} determinants for HF orbitals and for less than \num{5e5} determinants for natural orbitals,
a relatively small number of determinants.
We obtained $\Re[\Eexfci]$ values of $-109.326(2)$, $-109.325(1)$, $-109.324(1)$, and $-109.32370(3)$ \si{\hartree},
for HFN, HFA, NON, and NOA, respectively.
Despite the slight differences in the extrapolated values, they all overlap when accounting for the extrapolation errors.
Clearly, natural orbitals provide smaller extrapolation errors than HF orbitals.
Moreover, the extrapolation is less sensitive to the number of fitting points, as shown in the \SupMat.
Among the four sets of orbitals, the NOA comes out to be the best choice by a significant margin;
it has the fastest convergence in terms of the number of determinants, in addition to the smallest uncertainty associated with the extrapolation.

Concerning $\Im[\Esta]$, its evolution as a function of the number of determinants is depicted in Fig.~\ref{fig:mos_im_e_f_ndet}.
Contrary to $\Re[\Esta]$, it evolves more erratically before the stationary space reaches approximately \num{1e6} determinants.
Beyond this point, it converges smoothly from above with natural orbitals and from below with HF orbitals.
However, from the different behaviors, it is hard to conclude which set has the best rate of convergence.
For that, it is necessary to look at the evolution of $\Im[\Esta]$ as a function of $\Im[\Eapt]$, which is reported in Fig.~\ref{fig:mos_im_e_f_apt2}.
Only the curve associated with NOA exhibits a clear linear behavior,
whereas the three other curves would probably need a few additional CAP-CIPSI iterations to reach the linear regime.
Therefore, these three sets of orbitals should be less trustworthy for a linear extrapolation of $\Im[\Esta]$ to its FCI limit.
Indeed, the values of $\Im[\Eexfci]$ obtained from calculations employing HFN, HFA, NON, and NOA are, respectively,
$-0.007(1)$, $-0.007(1)$, $-0.0072(4)$, and $-0.00728(6)$ \si{\hartree}.
While all values are consistent among themselves,
the latest one has the smallest uncertainty and the least dependence on the number of fitting points (see \SupMat).
Therefore, we can clearly see that, again, it is more suitable to consider NOA orbitals.

When employing $\Im[\Ept]$ instead of $\Im[\Eapt]$, the curves exhibit an overall more linear behavior (as shown in the \SupMat).
However, as mentioned before,
this may be a problematic extrapolation criterion
because its individual components, $\Im[e_\alpha^{(2)}]$, can be either positive or negative.
Still, the extrapolated values are also consistent with those obtained with $\Im[\Eapt]$.


After establishing that NOA leads to the fastest convergence,
we now address the role of the selection procedure.
As described before,
we employed three different criteria for the selection of the determinants,
based on the norm of the perturbative correction ($\abs*{e_\alpha^{(2)}}$),
its real ($\Re[e_\alpha^{(2)}]$), or its imaginary ($\Im[e_\alpha^{(2)}]$) components.

\begin{figure*}
    \centering
    \begin{subfigure}[b]{0.475\textwidth}
        \centering
        \includegraphics[width=\textwidth]{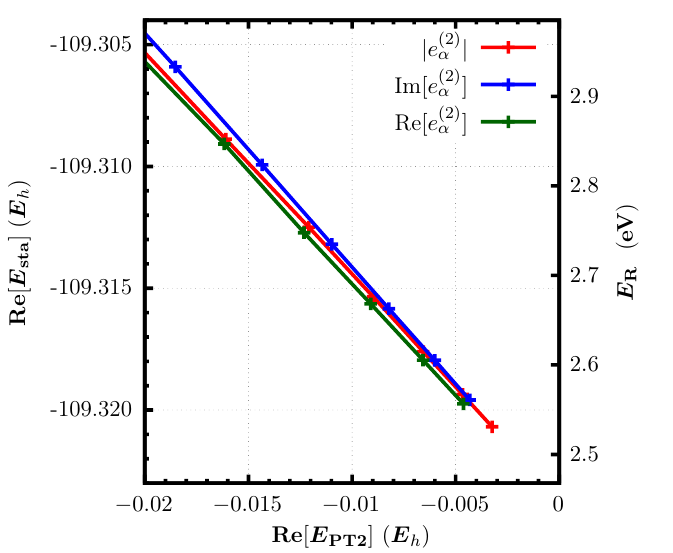}
        \caption{}
        \label{fig:crit_re_e_f_pt2}
    \end{subfigure}
    \hfill
    \begin{subfigure}[b]{0.475\textwidth}
        \centering
        \includegraphics[width=\textwidth]{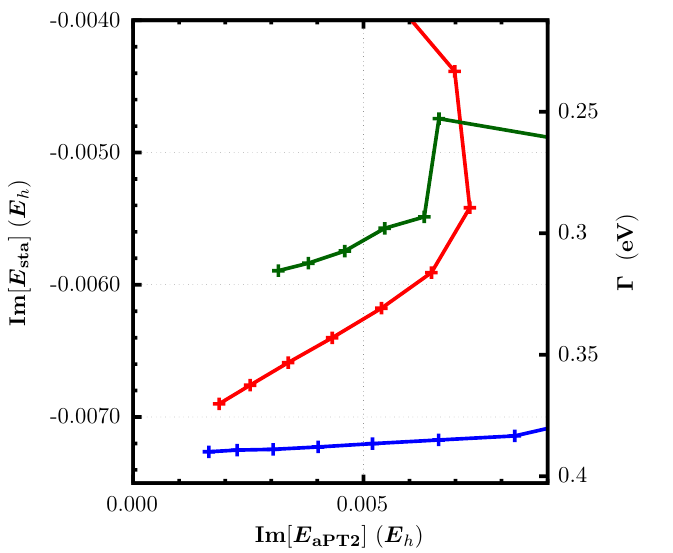}
        \caption{}
        \label{fig:crit_im_e_f_apt2}
    \end{subfigure}
	\caption{Evolution of the real [panel (a)] and imaginary [panel (b)] parts of the CAP-CIPSI energy, $\Re[\Esta]$ and $\Im[\Esta]$,
	as functions of the real and imaginary second-order energy corrections, $\Re[\Ept]$ and $\Im[\Eapt]$, for the $^2\Pi_g$ shape resonance of \ce{N2-} with the parameters of Table \ref{tab:comp_det}. Three selection criteria are considered: $\abs*{e_\alpha^{(2)}}$, $\Re[e_\alpha^{(2)}]$, and $\Im[e_\alpha^{(2)}]$.
	Resonance position $E_R$ and width $\Gamma$ are obtained for the fixed extrapolated energy of the neutral system.
	}
    \label{fig:crit}
\end{figure*}

Starting with $\Re[\Esta]$ as a function of $\Re[\Ept]$, Fig.~\ref{fig:crit_re_e_f_pt2} shows that the curves
associated with the different selection criteria are rather close and exhibit a clear linear behavior.
$\Re[\Eexfci]$ is estimated as $-109.32370(3)$ \si{\hartree} according to $\abs*{e_\alpha^{(2)}}$
and as $-109.32374(8)$ \si{\hartree} according to $\Im[e_\alpha^{(2)}]$, which are consistent with each other.
However, the selection based on $\Re[e_\alpha^{(2)}]$ produces a slightly different value, of \SI{-109.3240(2)}{\hartree},
having greater uncertainty and lying outside of the extrapolation errors of the estimates obtained with the two alternative selection criteria.
Hence, the results given by either $\abs*{e_\alpha^{(2)}}$ or $\Im[e_\alpha^{(2)}]$ criteria seem to be more trustworthy,
with a preference for the former due to its smaller extrapolation error.

In contrast to the real part, the evolution of $\Im[\Esta]$ as a function of $\Im[\Eapt]$,
represented in Fig.~\ref{fig:crit_im_e_f_apt2}, strongly depends on the selection criterion.
Both $\abs*{e_\alpha^{(2)}}$ and $\Im[e_\alpha^{(2)}]$ attain a linear regime,
and they produce very similar extrapolated values of $-0.00728(6)$ and $-0.00729(4)$ \si{\hartree}, respectively.
With the $\Re[e_\alpha^{(2)}]$ criterion, however, a linear extrapolation would yield \SI{-0.0063(7)}{\hartree}, very off from the two previous values,
reflecting the fact that a linear regime has not been reached in this case.

Overall, both $\abs*{e_\alpha^{(2)}}$ and $\Im[e_\alpha^{(2)}]$ selection criteria are sensible choices,
whereas the criterion based on $\Re[e_\alpha^{(2)}]$ is not recommended.
We prefer to rely on $\abs*{e_\alpha^{(2)}}$ to obtain our final FCI estimates,
as it is arguably a more natural generalization of the standard CIPSI selection for complex-valued energies.

The above findings do not seem to depend on the choice of orbitals (results for HFA orbitals are reported in the \SupMat).
Similarly, the trends observed for the choice of orbitals and the selection criteria remain unchanged for the first-order corrected energies (also shown in the \SupMat).


Having defined our optimal computational protocol (NOA orbitals and the $\abs*{e_\alpha^{(2)}}$ selection criterion),
we are now in the position to compute FCI estimates of the resonance position and width for the shape resonances of \ce{N2-} and \ce{CO-},
labeled CAP-exFCI from here on.
The results are gathered in Table \ref{tab:res} and compared to the CAP-EOM-EA-CCSD values extracted from Ref.~\onlinecite{Zuev_2014}.
Experimental resonance parameters (for the equilibrium geometry) are also reproduced for these two prototypical systems,
which were obtained by fitting theoretical models to experiment. \cite{Berman_1983,Ehrhardt_1968,Zubek_1977,Zubek_1979}

\begin{table*}
\caption{
Position $E_R$ and width $\Gamma$ of the shape resonance of \ce{N2-} and \ce{CO-}, in \si{\eV}, computed at the zeroth-order and first-order CAP-EOM-EA-CCSD and CAP-exFCI levels with the parameters of Table \ref{tab:comp_det}.
Extrapolation errors associated with the CAP-exFCI values are given in parentheses.}
\label{tab:res}
\begin{ruledtabular}
\begin{tabular}{ldddd}
                                    & \mc{2}{c}{\ce{N2-}}                       & \mc{2}{c}{\ce{CO-}} \\
                                      \cline{2-3}                                 \cline{4-5}
                                    & \mc{1}{c}{$E_R$}    & \mc{1}{c}{$\Gamma$} & \mc{1}{c}{$E_R$} & \mc{1}{c}{$\Gamma$} \\
\hline
Zeroth-order CAP-EOM-EA-CCSD\fnm[1] & 2.487               & 0.417               & 2.088            & 0.650               \\
Zeroth-order CAP-exFCI\fnm[2]       & 2.449(1)            & 0.391(3)            & 2.060(8)         & 0.611(3)             \\
Zeroth-order CAP-exFCI + basis set correction       & 2.470(1)\fnm[3]            & 0.338(3)\fnm[3]            & 1.898(3)\fnm[4]         & 0.765(3)\fnm[4]             \\
\\
First-order  CAP-EOM-EA-CCSD\fnm[1] & 2.571               & 0.255               & 1.981            & 0.585               \\
First-order  CAP-exFCI\fnm[2]       & 2.435(6)            & 0.31(1)             & 2.035(3)         & 0.696(5)             \\
First-order  CAP-exFCI + basis set correction       & 2.342(6)\fnm[5]            & 0.34(1)\fnm[5]             & 1.816(5)\fnm[6]          & 0.715(5)\fnm[6]             \\
\\
Experiment\fnm[7]                   & 2.316               & 0.414               & 1.50             & 0.75                \\
                                    &                     &                     & 1.52             & 0.80                \\
\end{tabular}
\end{ruledtabular}
\fnt[1]{Values taken from Ref.~\onlinecite{Zuev_2014}.}
\fnt[2]{This work.}
\fnt[3]{Basis set correction computed as the difference between the zeroth-order CAP-EOM-EA-CCSD values in the aug-cc-pVQZ+3s3p3d and aug-cc-pVTZ+3s3p3d basis set taken from Ref.~\onlinecite{Zuev_2014}.}
\fnt[4]{Basis set correction computed as the difference between the zeroth-order CAP-EOM-EA-CCSD values in the aug-cc-pV5Z+3s3p3d and aug-cc-pVTZ+3s3p3d basis set taken from Ref.~\onlinecite{Zuev_2014}.}
\fnt[5]{Basis set correction computed as the difference between the first-order CAP-EOM-EA-CCSD values in the aug-cc-pVQZ+3s3p3d and aug-cc-pVTZ+3s3p3d basis set taken from Ref.~\onlinecite{Zuev_2014}.}
\fnt[6]{Basis set correction computed as the difference between the first-order CAP-EOM-EA-CCSD values in the aug-cc-pV5Z+3s3p3d and aug-cc-pVTZ+3s3p3d basis set taken from Ref.~\onlinecite{Zuev_2014}.}
\fnt[7]{Experimental values for \ce{N2-} extracted from Ref.~\onlinecite{Berman_1983}, and for \ce{CO-} extracted from Refs.~\onlinecite{Ehrhardt_1968,Zubek_1977,Zubek_1979}.
}
\end{table*}

Our CAP-exFCI resonance parameters can be considered chemically accurate for the given basis set and CAP parameters.
The uncertainties for the resonance positions are \SI{0.001}{\eV} (\ce{N2-}) and \SI{0.008}{\eV} (\ce{CO-}),
whereas for the widths they are \SI{0.003}{\eV} for both systems.
Larger uncertainties arise for the first-order corrected energies [see Eq.~\eqref{eq:E_tilde}], though still below \SI{0.01}{\eV}.
This is understandable as the first-order correction changes the stationary energy $\Esta$ but not the perturbative energy correction $\Ept$,
thus rendering the extrapolation curves less linear than those involving the zeroth-order stationary energy $\Esta$ (see the \SupMat).

For \ce{N2-}, we find that zeroth-order CAP-EOM-EA-CCSD delivers very close results to zeroth-order CAP-exFCI,
with slightly overestimated resonance position and width, by $0.038(1)$ and \SI{0.026(3)}{\eV}, respectively.
The discrepancy for the resonance position is consistent with the typical errors of EOM-CCSD for excitation energies of bound states. \cite{Veril_2021}
We also find milder first-order corrections with CAP-exFCI [resonance position changes by \SI{-0.013(6)}{\eV} and width by \SI{-0.09(2)}{\eV}]
than with CAP-EOM-EA-CCSD (resonance position changes by \SI{+0.084}{\eV} and width by \SI{-0.162}{\eV}).
This in turn deteriorates the favorable comparison observed for the zeroth-order results.
The first-order CAP-EOM-EA-CCSD resonance position becomes even more overestimated, by \SI{0.136(6)}{\eV},
whereas the width now appears underestimated, by \SI{-0.05(1)}{\eV},
with respect to first-order CAP-exFCI.

The findings for \ce{CO-} are overall similar to those discussed for \ce{N2-}.
CAP-EOM-EA-CCSD compares more favorably with CAP-exFCI in their zeroth-order versions
[resonance positions and width slightly overestimated, by $0.028(8)$ and \SI{0.039(3)}{\eV}]
than their first-order counterparts
[resonance positions and width underestimated, by $-0.058(3)$ and \SI{-0.111(5)}{\eV}].
Similarly, the first-order correction has a less pronounced effect on the resonance position obtained with CAP-exFCI [\SI{-0.025(6)}{\eV}] than with CAP-EOM-EA-CCSD (\SI{-0.107}{\eV}).

Accounting for resonance positions and widths of both systems,
zeroth-order CAP-EOM-EA-CCSD has a mean absolute error of \SI{0.033(4)}{\eV} with respect to CAP-exFCI,
which increases to \SI{0.087(9)}{\eV} when comparing the first-order corrected methods.
These results suggest that the apparent good performance of zeroth-order CAP-EOM-EA-CCSD is partially due to error cancelation
stemming from the presence of the CAP and from the method's restriction to double excitations.

When comparing to experiment, it is clear that correlation effects captured at the CAP-exFCI level have a major impact.
Taking the more reliable first-order corrected results,
going from CAP-EOM-EA-CCSD to CAP-exFCI significantly reduces the gap with respect to experiment
for the resonance position of \ce{N2-} [from \SI{0.255}{\eV} to \SI{0.119(6)}{\eV}],
and for the resonance widths of \ce{N2-} [from \SI{0.159}{\eV} to \SI{0.108(1)}{\eV}] and \ce{CO-} [from \SI{0.19}{\eV} to \SI{0.079(5)}{\eV}].
Furthermore, by looking into relative differences instead of absolute differences,
we see that higher-order correlation effects are more pronounced for the resonance widths than for the resonance positions.
The remaining differences to experiment should be related to basis set effects, known to be particularly relevant for these shape resonances, \cite{Zuev_2014,Falcetta_2014}
and the CAP itself, with an associated error that remains less understood.

\alert{We can estimate the basis set effect from
the difference between the CAP-EOM-EA-CCSD values obtained in a more complete basis set (aug-cc-pVQZ+3s3p3d for \ce{N2-} and aug-cc-pV5Z+3s3p3d for \ce{CO-})
and in the aug-cc-pVTZ+3s3p3d basis set, all extracted from Ref.~\onlinecite{Zuev_2014}.
Whereas such basis set correction based on lower-level computational models is a common practice for bound state calculation,
it has not yet been carefully studied in the case of resonances.
That being said, the correction brings the first-order CAP-exFCI results closer to experiment.
As shown in Table \ref{tab:res}, the effect is modest for the resonance widths
but is more significant for the resonance positions, where a close match with experiment is seen for \ce{N2-}.
The basis set correction also improves the comparison with the experimental resonance position of \ce{CO-}, but the gap remains substantial nonetheless.
While this could be related to the error introduced by the CAP, the slow basis set convergence suggests that even larger basis sets would be needed.
Because of that, the present basis set correction also becomes more questionable.
Assuming there is no major inaccuracies in the experimental values,
it is interesting to notice that \ce{CO-} remains more challenging for theory than \ce{N2-},
despite the qualitatively similar character of the two isoelectronic shape resonances.}


In conclusion, we have reported the first implementation of SCI for resonances using the CAP technique.
As a first application, we have shown that the resulting CAP-SCI methodology allows us to produce FCI estimates
of the position and width of the shape resonances of two paradigmatic transient anions, \ce{N2-} and \ce{CO-}.

To reach this level of accuracy, the choice of the orbitals plays a critical role.
We have found that (real-valued) natural orbitals obtained specifically for the resonant state are particularly well-suited.
There is however room for improvement and one could employ other kinds of orbitals.
Among them, we can mention complex-valued natural or energetically-optimized orbitals,\cite{Levine_2020,Yao_2021,Damour_2021} although it would imply significant modifications of the current CIPSI algorithm.
The way to select the determinants has also been found to be crucial as they must be selected based on both the real and the imaginary parts of the stationary energy.

Our results indicate that, for a given set of CAP parameters and basis set,
the higher-order correlation effects fully accounted for in the CAP-SCI methodology can explain up to half of the difference between CAP-EOM-EA-CCSD results \cite{Zuev_2014} and experiment.
The remaining deviation between theory and experiment can thus be attributed to the finite basis set error and/or the approximate treatment of the continuum by the CAP.
We also discussed how zeroth-order CAP-EOM-EA-CCSD is probably subject to partial error cancelation.

As a perspective, we are planning on extending SCI to other adapted quantum chemistry methodologies, such as complex scaling and complex basis functions.
We are also interested in studying resonances where SCI is expected to be particularly well adapted.
As an example, the two-particle one-hole Feshbach resonances of water and benzene are known to be challenging systems for state-of-the-art methodologies. \cite{Thodika_2020,Thodika_2022}
We hope to report on this in the near future.

\acknowledgements{
This project has received funding from the European Research Council (ERC) under the European Union's Horizon 2020 research and innovation programme (Grant agreement No.~863481). This work used the HPC resources from CALMIP (Toulouse) under allocation 2024-18005.}

\section*{Supporting Information}
\label{sec:supmat}
See the \SupMat for the parameters of the additional basis functions, the description of the extrapolation procedure, as well as additional tables and figures regarding the influence of the orbital set, the selection criterion, and analogous results for the first-order corrected results.

\section*{References}

\bibliography{cap_cipsi}

\end{document}